\documentclass[aps]{revtex4-1}
\usepackage{graphicx}
\usepackage{caption}
\usepackage{subcaption}
\usepackage{dcolumn}
\usepackage{bm}
\usepackage{amsmath}
\usepackage{amssymb}
\usepackage{epsfig}
\usepackage{verbatim}
\usepackage{pstricks,pst-grad,pst-plot}
\usepackage{natbib}
\usepackage{url}
\usepackage{makeidx} 
\usepackage{wrapfig} 
\usepackage{float} 

\begin{document} 
\title{Effects of variable neighbourhood on spreading processes}
\author{Sergei~N.~Taraskin}
\affiliation{St. Catharine's College and Department of Chemistry,
University of Cambridge, Cambridge, UK}
\email{snt1000@cam.ac.uk}
\author{Francisco~J.~P{\'e}rez-Reche}
\affiliation{Institute for Complex Systems and Mathematical Biology, SUPA, University of Aberdeen, Aberdeen, UK }
\email{fperez-reche@abdn.ac.uk}

\begin{abstract}
A theoretical framework for the description of Susceptible-Infected-Removed (SIR) 
spreading processes with synergistic transmission of infection on a lattice is developed. 
The model incorporates explicitly the effects of a time-dependent environment on the transmission of infection between hosts. 
Exact solution of the model shows that time-dependence of the neighbourhood of recipient hosts is a key factor for synergistic spreading processes.
It is demonstrated that the higher the connectivity of a lattice, the more prominent is the effect of synergy on spread.
\end{abstract}

\pacs{87.23.Cc, 05.70.Jk, 64.60.De, 89.75.Fb}

\maketitle
\date{\today}
 
\section{\label{sec:Introduction}Introduction} 

Spreading of infection, opinion and rumours through networks of hosts is a topic of very active interdisciplinary  research~\cite{Vespignani_NaturePhys2012,Castellano_RMP2009}. 
The standard models for spreading processes such as the prototype susceptible-infected-removed (SIR) model~\cite{grassberger1983} assume independent uncorrelated transmission of the spreading agent between hosts. 
However, non-linear synergistic effects caused by interactions between hosts are becoming increasingly recognised as key factors for the spread of behaviour~\cite{Centola_Science2010}, opinion dynamics~\cite{Castellano_RMP2009} and  transmission of pathogens~\cite{Ben-Jacob_Nature1994,Ludlam_2012}.  

Synergistic effects of the type studied in this paper can be illustrated using the spread of opinion as a benchmark.
Consider a situation when a host $A$ tries to transmit an opinion to a host $X$ which is surrounded by hosts $B$ and $C$ in addition to $A$.  
If the hosts $B$ and $C$ do not influence the process of transmission, this occurs with probability $P(A\to X)$. 
However, hosts $B$ and $C$ will typically influence the transmission of opinion from $A$ to $X$ either supporting (positive synergy) or interfering with (negative synergy) transmission. 
In this case, host $X$ will adopt the opinion of $A$ with a probability $P(A\to X|B,C)$ that can differ from $P(A\to X)$. 
These synergistic effects on individual level can significantly change the behaviour of the spreading process on a global scale, e.g. make it invasive or {\it vice versa} can lead to its localisation. 

Several theoretical models have been developed to incorporate synergistic effects in the description of spreading processes~\cite{Dodds_PRL2004,Centola_PhysicaA2007,Castellano_RMP2009,Perez_Reche_2011:PRL,Lu_NewJPhys2011,Krapivsky_JSTAT2011,Bizhani_PRE2012,ZhengLuMing_PRE2013}. 
In particular, the effects of constructive and destructive synergy caused by the nearest neighbours of the pair of hosts transmitting infection in between were considered for SIR process in Ref.~\cite{Perez_Reche_2011:PRL}. 
The synergistic model presented in that paper has been extensively analysed numerically and within a simple analytical toy model that reproduces the numerical results at a qualitative level. 
This toy model incorporates the dependence of the transmission rates on the neighbourhood which is an essential feature of synergistic spreading processes. 
However, it neglects several factors that do not play any role in the non-synergistic SIR processes but are expected to be essential in the presence of synergy.
One of the neglected factors is a possible variation of the neighbourhood of a pair of hosts (involved in transmission) with time. 
Namely, it was assumed in the toy model that if at initial moment of time when the host $A$ started challenging host $X$, the host $B$ supported the spread and the host $C$ did not influence the process, then 
neither $B$ nor $C$ change their state (supportive and neutral, respectively) during the fixed duration of transmission. 
In fact, as shown below, this assumption is equivalent to the assumption of synchronous dynamics in transmission.
Bearing in mind the stochastic nature of transmission, it is unlikely   
that hosts $B$ and $C$ do not change their state during transmission between 
$A$ and $X$, i.e. dynamics of transmission is likely to be asynchronous.
By definition, synergistic transmission depends on the evolution of the neighbourhood. 
Specifically, the probability that $X$ is challenged by two or more neighbours during its lifetime and synchrony or asynchrony between challenging neighbours are expected to be key factors affecting transmission. 
For instance, it is intuitively clear that if hosts $A$, $B$ and $C$ are synchronised to start challenging $X$ at the same time, synergistic effects will be enhanced with respect to a case with de-synchronised action.   
Another assumption used in the synergistic toy model of Ref.~\cite{Perez_Reche_2011:PRL} is that the probabilities 
of various neighbourhoods (e.g. host $B$ is supportive and host $C$ is neutral) do not depend on transmission rates. 
This assumption clearly breaks in the limiting cases of e.g. very high transmission rates when both the hosts $B$ and $C$ are very likely to share the opinion and thus both be supportive. 

In this paper, we relax these two simplifying assumptions,  
develop and solve an analytical model for the SIR process which accounts for the time-dependent nearest-neighbour environment and its influence on transmission of infection. 
We show that, in contrast to non-synergistic SIR processes~\cite{Ludwig_MBiosc1975,PerezReche_Interface2012}, incorporating time-dependence of the neighbourhood is essential to properly account for key epidemiological quantities such as the probability of invasion. 
The effects of synergy on the epidemic threshold for transmission rates are analysed in several two-dimensional (2d) lattices characterised by different number of nearest neighbours. 

The structure of the paper is the following. 
The model is introduced in Sec.~\ref{sec:Model} and solved in Sec.~\ref{sec:Solution} for asynchronous transmission dynamics. 
The analytical results are presented and compared with the results of numerical
simulations for several 2d lattices in Sec.~\ref{sec:Results}. 
The effect of synchrony in transmission of infection is briefly discussed in Sec.~\ref{sec:Synchrony}.
The conclusions are given in Sec.~\ref{sec:Conclusions}. 
Some technical details are discussed in App.~\ref{app:A} and App.~\ref{app:B}.

\section{\label{sec:Model}Model}

Let us consider the conventional SIR process on a regular lattice with coordination number (node degree) $q$. 
Assume that an arbitrary susceptible (S) node $i$ has been infected at time $t=0$ by one of its infected (I) neighbours. 
It remains in the infected state in the time interval $[0,\tau_i]$ and then becomes removed (R). 
This node attempts to transmit the infection during its lifetime in state I to one of the susceptible nearest neighbours, $j$. 
The process of transmission of the infection is of the Poisson type and occurs with probability, $T(t)$,
\begin{eqnarray} 
 T(t) = 1-e^{-\int_0^t \lambda(t) \text{d}t}~,  
\label{eq:T_t} 
\end{eqnarray} 
where $T(t)$ is the probability for node $i$ to transmit infection to node $j$ by time $t$ and $\lambda(t)$ is the  transmission rate that can, in principle, depend on the time passed since the moment of infection of the donor host, $i$. 
The probability, $T(\tau_i)$, to transmit the infection from node $i$ to node $j$ during the life-time of node $i$, $\tau_i$, is the transmissibility along $i-j$ link. 

Transmissibilities are key quantities determining whether an SIR epidemic is invasive (i.e. spans the system when the transmissibilities are large enough) or non-invasive (i.e. does not span the system when the transmissibilities are relatively small). 
In general, the invasion threshold separating the invasive and non-invasive regimes depends on the whole set of transmissibilities between all the pairs of hosts. 
However, simpler descriptions are possible under certain conditions. 
For example, in a homogeneous non-synergistic system with identical transmissibility $T$ between all connected hosts, the SIR epidemic can be mapped onto uncorrelated dynamical percolation~\cite{grassberger1983} and the value of $T$ at the invasion threshold coincides with the bond-percolation threshold, $T_c$, for the corresponding network.
In systems with heterogeneous but uncorrelated transmissibilities, the final state of an epidemic is fully characterised by the mean value of transmissibility, $\langle T \rangle$, defined as the average over all pairs of connected hosts for a particular realisation of epidemic and then over different realisations of epidemics. 
In this case, the mapping to uncorrelated dynamical percolation~\cite{grassberger1983} with bond probability still holds and the invasion threshold is given by the following equation~\cite{Cox1988,sander2002,PerezReche_JRSInterface2010}:
\begin{eqnarray} 
\langle T \rangle = T_c~.
\label{eq:condition} 
\end{eqnarray} 
In case of correlations between transmissibilities, resulting e.g. from synergy  between hosts, the final state of an epidemic is not completely determined by the value of $\langle T \rangle $~\cite{kuulasmaa1982,Kenah2007,Miller_JApplProbab2008,Neri_JRSInterface2011}. 
However, Eq.~\eqref{eq:condition} still provides reasonable estimates of the invasion threshold in a wide range of situations, as demonstrated below.

If the transmission of the infection can be described by a homogeneous (in time) Poisson process, then the transmission rate $\lambda(t)$ does not vary with time and $\lambda(t)=\alpha$. 
However, in reality, the transmission rate can vary with time and transmission becomes a non-homogeneous Poisson process. 
A possible reason for such time dependence can be due to the influence on transmission from the nearest neighbours of the node under attack, i.e. due to the synergy effects~\cite{Perez_Reche_2011:PRL}. 
In this case, the transmission rate $\lambda(t)=0$ for $t\in (-\infty,0)\cup(\tau_i,\infty)$ and 
\begin{subequations}
\label{eq:lambda_t}
\begin{equation} 
 \lambda(t) = \alpha + n(t) \beta ~~~\text{if}~~~\alpha +n(t) \beta>0~~~~\text{and} 
\label{eq:lambda_t_a}
\end{equation}
\begin{equation}
\lambda(t)= 0 ~~~\text{if}~~~\alpha +n(t)\beta\le 0~,  
\label{eq:lambda_t_b} 
\end{equation}
\end{subequations}
for $t\in [0,\tau_i]$. 
Here, in the case of so-called recipient$(r)-$synergy~\cite{Perez_Reche_2011:PRL}, the value of $n(t)$ is the number of infected nearest neighbours of node $j$ in addition to the attacking node $i$ ($0\le n(t) \le q-1$), $\alpha$ is the synergy-free transmission rate and $\beta$ is the synergy contribution to the transmission rate (synergy rate). 
In what follows, except for the discussion in Sec.~\ref{sec:Synchrony}, both rates, $\alpha$ and $\beta$, are assumed to be time-independent. 
Positive values of $\beta$ describe the constructive synergy effects when the presence of infected neighbours of a node $j$ helps to transmit infection from attacking node $i$ to $j$. 
Negative values of $\beta$ refer to interference in the transmission when infected neighbours $j'$ of $j$ obstruct the transmission of infection from node $i$ to $j$. 
The number of infected  nearest neighbours of node $j$ can vary with time and thus the transmission of infection from $i$ to $j$ can occur in a time-dependent environment. 

In the simplest approximation used in~\cite{Perez_Reche_2011:PRL}, the number of infected nearest neighbours does not change with time and is equal to the number of infected nearest neighbours of node $j$ at the moment of infection of node $i$.  
Below, we relax this simplification and take into account the effects of a time-dependent neighbourhood on transmission directly. 

If the life-time of all nodes in infected state is the same, $\tau_i=\tau$, then the nearest neighbours $j'$ of node $j$ can affect the transmission of infection from $i$ to $j$, only if they are infected at times $t_{j'}$ in the time interval, 
$t_{j'}\in [-\tau,\tau]$ (below, the time-scale is defined by setting up $\tau=1$). 
In what follows, we assume, for simplicity and in consistency with Eq.~\eqref{eq:condition}, that the infection times, $t_{j'}$, of nodes $j'$ and node $i$ are independent and identically distributed random variables with 
even probability density function,  $\rho(t_{j'})=\rho(-t_{j'})$. 
The probability $s=s(\alpha,\beta)$ for node $j'$ to be infected within the time interval $t_{j'}\in [-1,1]$, and thus 
being able to affect the transmission of  infection  from node $i$ to $j$ is given by,
\begin{equation}
s=\int_{-1}^1 \rho(t)\text{d}t~.
\end{equation} 

The mean value of transmissibility between nodes $i$ and $j$ can be calculated 
by accounting for all possible synergistic contributions from nearest neighbours $j'$ ($j'\ne i$) of node $j$, i.e. 
\begin{eqnarray} 
\langle T \rangle &=& \int\limits_{-\infty}^\infty\ldots \int\limits_{-\infty}^\infty 
\left( 1 - \exp{\left[-r(t_1,\ldots,t_{n})
 \right]} \right)
 \prod_{j'=1}^{q-1}\rho(t_{j'}) \text{d}\, t_{j'}
\nonumber \\
 &=& \sum_0^{q-1} \binom{q-1}{n} (1-s)^{q-1-n} P_n 
~.
\label{eq:T_mean_general} 
\end{eqnarray} 
Here, 
\begin{equation}
r(t_1,\ldots,t_{n})= \int_0^{1} \lambda(t;t_1,\ldots,t_{n})\text{d}\, t~,
\end{equation}
and the value of $P_n$ given by the following expression, 
\begin{eqnarray}  
P_n &=&\int\ldots\int\limits_{C_{n}}
\left( 1 - \exp{\left[-\int_0^{1} \lambda(t;t_1,\ldots,t_{n})\text{d}\, t
 \right]}\right) \prod_{j'=1}^{n}\rho(t_{j'}) \text{d}\, t_{j'} 
~,
\label{eq:P_n} 
\end{eqnarray} 
is the probability of passing the infection from node $i$ to node $j$ in the presence of $n$ particular nearest neighbours affecting the infection, i.e. exhibiting r-synergy.
The integration in Eq.~\eqref{eq:P_n} is performed over the interior domain of an $n$-dimensional cube, $C_{n}$, with side-length $=2$, i.e. $C_{n}: t_{j'}\in[-1,1] (j'=1,\ldots,n)$. 
In case of no affecting nearest neighbours, $n=0$,  the synergy effects are absent and 
\begin{eqnarray}  
P_0=1-e^{-\alpha}
~.
\label{eq:P_0} 
\end{eqnarray} 
Similarly, if $\beta=0$ the mean transmissibility coincides with $P_0$. 

Eqs.~\eqref{eq:T_mean_general}-\eqref{eq:P_n} give the expression for the mean transmissibility in the case of a time-dependent environment for transmission of infection in the presence of synergistic effects. 
Technically, the calculation of $\langle T \rangle $ is reduced to the evaluation of $n$-dimensional integrals $P_n$ which can be done, as demonstrated below, in closed forms. 

\section{\label{sec:Solution}Solution for asynchronous transmission}

In this Section, we assume that $\rho(t_{j'})$ does not change significantly on the time-scale of  $\tau=1$ and use the following uniform distribution:
\begin{eqnarray} 
\rho(t_{j'})\simeq \frac{s}{2}~, ~~~\text{for}~~~ t_{j'}\in [-1,1]
~.
\label{eq:rho_uniform} 
\end{eqnarray} 
This is expected to be a reasonable approximation for asynchronous transmission dynamics (the effect of synchrony is discussed in Sec.~\ref{sec:Synchrony}).
The dependence of $s=s(\alpha,\beta)$ (the only parameter of the model) on synergy-free and synergy rates can be found numerically from the simulations 
for the SIR process with synergy (see below). 
In the toy model proposed in Ref.~\cite{Perez_Reche_2011:PRL}, 
the value of $s$ was assumed to be independent of $\alpha$ and $\beta$. 
Under assumption \eqref{eq:rho_uniform}, the probability $P_n$ can be expressed as $P_n= (s/2)^n I_n$ in terms of the integrals,
\begin{equation}
\label{eq:In}
I_n=\int\ldots\int\limits_{C_{n}}
\left( 1 - \exp{\left[-r(t_1,\ldots,t_{n})
 \right]}\right) \prod_{j'=1}^{n} \text{d}\, t_{j'}
\end{equation}

Below, we derive an exact expression for the mean transmissibility by means of direct evaluation of the integrals  $I_n$ for arbitrary value of $n$. 
The complications in their evaluation come from  condition~\eqref{eq:lambda_t_b} leading for $q>2$ to several distinct intervals for negative $\beta$, in which the integrals, $I_n$, have different form. 
We will start with a particularly straightforward case of relatively large values of $\beta$ when the negative synergy cannot reduce the transmission rate to zero. 
In this case, it is relatively straightforward to deal directly with $P_n$ and obtain expressions valid for generic $\rho(t_{j'})$ that will also be used in Sec.~\ref{sec:Synchrony}.

\subsection{\label{sec:sol_1}Exact result for mean transmissibility in the regime  $\beta \ge -\alpha/(q-1)$}

In the regime of relatively large transmission  rates, $\beta \ge  -\alpha/(q-1)$, the time-dependent transmission rate is given by 
Eq.~\eqref{eq:lambda_t_a} and can never be equal to zero for $t\in [0,1]$. 
The integral $r(t_1,\ldots,t_n)$ in the integrand of $P_n$ (see Eq.~\eqref{eq:P_n}) can be evaluated  for an arbitrary number $n$, $1\le n \le q-1$, of nearest neighbours infected at times $t_{j'}$ 
with $t_{j'}\in [-1,1]$, $j=1,\ldots,n$, as follows, 
\begin{eqnarray}
r(t_1,\ldots,t_{n})=  \alpha +n\beta - \beta \sum_{j'=1}^{n}|t_{j'}|~.
\label{eq:r}
\end{eqnarray}

Substitution of Eq.~\eqref{eq:r}  into the expression for  $P_n$ given by Eq.~\eqref{eq:P_n}  
 results in 
%
%
\begin{eqnarray} 
P_n&=& \int\limits_{-1}^1\ldots \int\limits_{-1}^1
\left( 1 - e^{-(\alpha+n\beta)}\prod_{j'}^{n}e^{\beta |t_{j'}|}\right)\prod_{j'}^{n}  \rho(t_{j'})\text{d}\, t_{j'}
\nonumber \\
&=& s^n-e^{-\alpha}\left(sf(\beta)\right)^n
~,
\label{eq:T_mean_cube} 
\end{eqnarray}
where the function,
\begin{equation}
f(\beta)\equiv \frac{2e^{-\beta}}{s} \int_0^1 e^{\beta t} \rho(t) \text{d}t~,
\label{eq:fbeta_def}
\end{equation}
takes the form,
\begin{equation}
f_0(\beta)=\frac{1-e^{-\beta}}{\beta}~,
\label{eq:fbeta_uniform}
\end{equation}
for uniform $\rho(t_{j'})$ (see Eq.~\eqref{eq:rho_uniform}).

Substitution of Eq.~\eqref{eq:T_mean_cube} into Eq.~\eqref{eq:T_mean_general} leads to the following expression for the mean transmissibility, 
\begin{eqnarray}
\langle T \rangle &=& 1 - e^{-\alpha}\sum_{n=0}^{q-1} \binom{q-1}{n} (1-s)^{q-1-n}s^n
\left(f(\beta) \right)^{n} 
\nonumber \\
&=& 1 - e^{-\alpha}\left(1-s +s f(\beta) \right)^{q-1}
~. 
\label{eq:T_mean_total} 
\end{eqnarray} 
in the case of time-dependent neighbourhood for $\beta > -\alpha/(q-1)$. 

The phase boundary, $\alpha_c(\beta)$, in this region of the $(\alpha,\beta)$  plane, i.e. for $\beta > -\alpha/(q-1)$, can be obtained approximately from 
Eq.~\eqref{eq:T_mean_total} by replacing $\langle T \rangle$ with 
the bond-percolation threshold, $T_{\text{c}}$, in the form of an implicit 
equation for $\alpha_c(\beta)$, 
\begin{eqnarray}
T_c
=  1 - e^{-\alpha_c(\beta)}\left(1-s(\alpha_c(\beta),\beta) +s(\alpha_c(\beta),\beta) f(\beta) \right)^{q-1}
~,  
\label{eq:boundary_1} 
\end{eqnarray} 
which can be solved numerically for given $f(\beta)$ once the function $s(\alpha,\beta)$  is known. 

In the limiting case of the small synergy rate, $\beta \to 0$, Eq.~\eqref{eq:boundary_1} can be solved approximately giving, 
\begin{eqnarray}
\alpha_c\simeq \alpha_0 +(q-1)s_0 f^{\prime}(0) \beta
~,
\label{eq:boundary_small_beta} 
\end{eqnarray} 
where $s_0=s(\alpha_0,0)$, $f^\prime(0)=-1/2$ for uniform $\rho(t_{j'})$ and $\alpha_0=-\ln(1-T_c)$ is the critical synergy-free transmission rate.
It follows from Eq.~\eqref{eq:boundary_small_beta} that the critical transmission rate $\alpha_c$ decreases with increasing synergy rate $\beta$ 
faster in networks with higher coordination number $q$ provided that the value of $s_0$ increases with $q$ (which is, actually, the case as shown below). 

The time-dependence of the environment has an important impact on the invasion threshold. Indeed, under assumption of a time-independent environment~\cite{Perez_Reche_2011:PRL}, the phase boundary is given 
by Eq.~\eqref{eq:boundary_1} with $f(\beta) = e^{-\beta}$   which results in a different asymptotic behaviour for large $\beta\to \infty$ and a twice larger gradient in the regime of small $\beta\to 0$ for uniform $\rho(t_{j'})$ since $f^\prime(0)=-1$ (see Eq.~\eqref{eq:boundary_small_beta}).

\subsection{\label{sec:sol_2}Exact result for mean transmissibility in the regime  $\beta < -\alpha$}

In the regime with $\beta \le -\alpha$, the synergistic interference  is so strong that the transmission rate equals zero when 
the recipient node, $j$, is challenged by more than one infected neighbour
 and $r=\alpha$ when none of them are infected. 
This means that the value of $r$ does not depend on $\beta$ and the integral $I_n(\alpha)$
in Eq.~\eqref{eq:P_n} can be evaluated exactly for arbitrary value of $n$. 
In fact, the following recursive relation can be established  for $I_n(\alpha)$, 
(see App.~\ref{app:A} for more detail), 
\begin{eqnarray}  
 I_n(\alpha)= 
(n+1)\left(1 -\frac{I_{n-1}(\alpha)}{\alpha} 
 \right)
~, ~~~n\ge 1  
\label{eq:I_n_neg_beta} 
\end{eqnarray} 
with $I_0=P_0$ (see Eq.~\eqref{eq:P_0}). 
The phase boundary, corresponding to $\alpha_c=\text{Const}$, can be found by solving the following implicit equation, 
\begin{eqnarray} 
T_c 
 = \sum_0^{q-1} \binom{q-1}{n} (1-s_{-})^{q-1-n}\left(\frac{s_{-}}{2}\right)^n I_n(\alpha) 
~,  
\label{eq:alpha_c_neg_beta} 
\end{eqnarray} 
with $I_n$ obeying Eq.~\eqref{eq:I_n_neg_beta} and $s_{-}=s(\alpha,-\alpha)$. 
For example, in case of $q=3$ (hexagonal lattice), this equation reduces to the following implicit equation for $\alpha=\alpha_c$,  
\begin{eqnarray} 
T_c
&=&
(1-s_{-})^2\left(1-e^{-\alpha}\right) +2s(1-s_{-})\left(1-\frac{1-e^{-\alpha}}{\alpha}\right) 
\nonumber \\
&+& 
3\left(\frac{s_{-}}{2}\right)^2 
\left[ 
1-
\frac{2}{\alpha}\left(1 - \frac{1-e^{-\alpha} }{\alpha} \right)
\right]
~, 
\label{eq:alpha_c_hex_neg_beta} 
\end{eqnarray} 
which can be solved numerically. 

\subsection{\label{sec:sol_3}Exact results for mean transmissibility in the regime  $-\alpha \le \beta < -\alpha/(q-1)$}

In the regime, $-\alpha \le \beta < -\alpha/(q-1)$, the integrals $I_n$ can be 
analytically evaluated separately in the following interval, $-\alpha/(n-1) \le \beta < -\alpha/n$, for $n=2,\ldots, q-1$ ($q\ge 3$). 
This results in monotonically decaying function $\alpha_c(\beta)$ with increasing $\beta$. 
The function is continuous but experiences  kink singularities 
(finite discontinuities in the first derivative)  at 
the boundaries of the regimes, i.e. when $\beta=-\alpha/n$,  $n=1,\ldots, q-1$.  
The functional form of  $I_n(\alpha,\beta)$ depends on $q$ and is different in each of the intervals. 
In App.~\ref{app:B}, we derive the results for the simplest case of $q=3$ for illustration. 

\section{\label{sec:Results} Results for asynchronous transmission}

In this section, we present the results for the phase diagrams in the $(\alpha,\beta)$ parameter space 
for the SIR process occurring on regular 2d-lattices of several topologies. 
The phase boundary $\alpha_c(\beta)$ separates the non-invasive 
($\alpha < \alpha_c(\beta)$) and invasive ($\alpha > \alpha_c(\beta)$) regimes of the SIR process. 
The numerical results for $\alpha_c(\beta)$  were obtained by simulating the  SIR process by using a continuous-time algorithm for asynchronous transmission dynamics and analysing the scaling behaviour of 1d-spanning epidemics in a similar way as in Ref.~\cite{Perez_Reche_2011:PRL}. 
Thus obtained phase boundaries served as tests for the analytical model described 
in the previous Section. 

\begin{figure}[h] 
\centering
\begin{subfigure}[b]{0.4\textwidth}
\centering
{\includegraphics[clip=true,width=\textwidth]{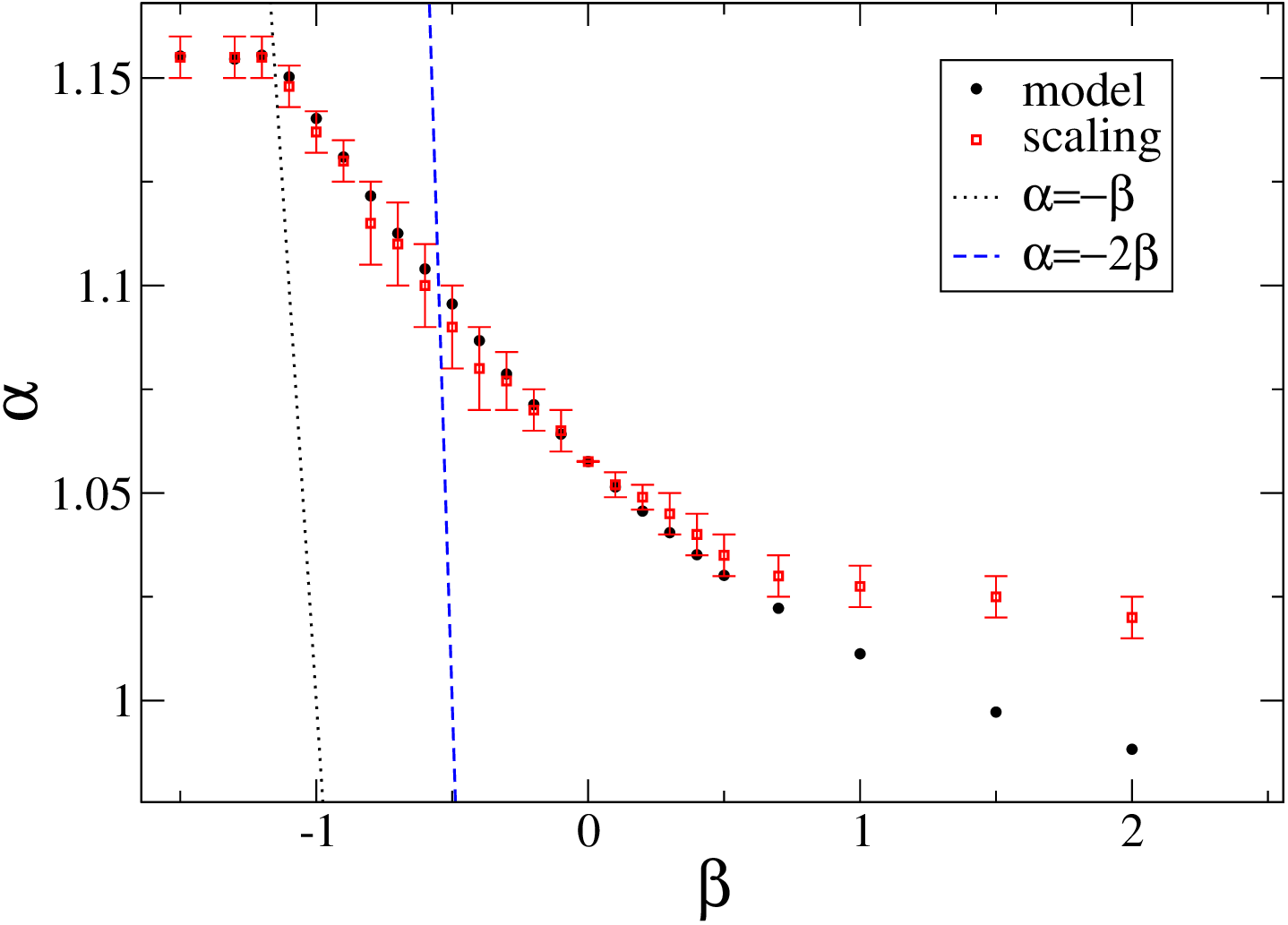}}   
\end{subfigure}
\quad
\begin{subfigure}[b]{0.4\textwidth}
\centering
{\includegraphics[clip=true,width=\textwidth]{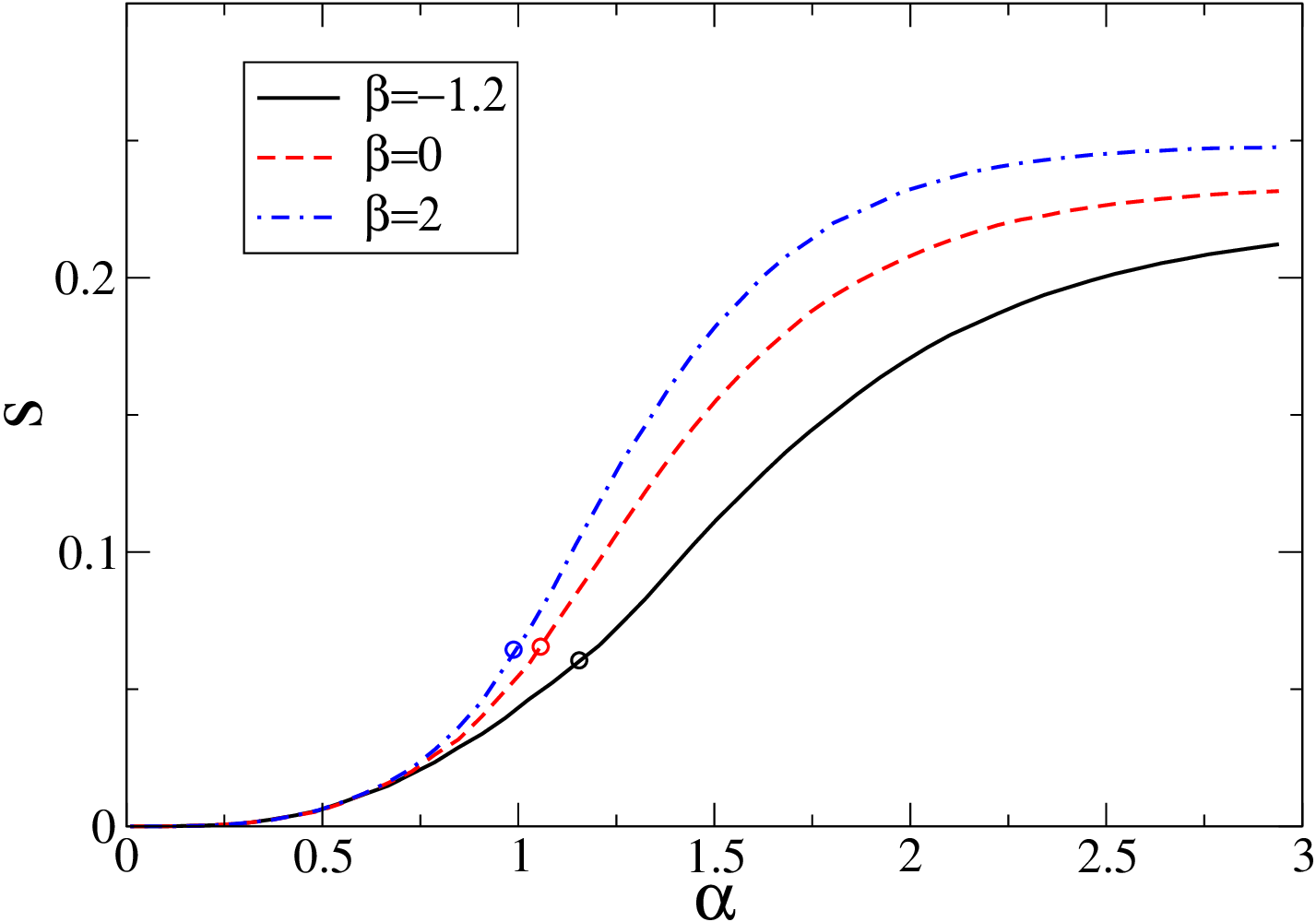}} 
\end{subfigure}
\caption{ (a) Phase diagram for hexagonal lattice. 
The results of numerical  analysis are shown by open symbols and were obtained 
by finite-size scaling for lattices of linear size $L$ ($N=L\times L$) with 
$L=30,\ldots,100$. The solid symbols correspond to the solution of the analytical model. 
The dotted and dashed lines correspond to $\alpha=-\beta$ and $\alpha=-2\beta$. \\
(b) Infection probability $s$ {\it vs} synergy-free transmission rate $ \alpha$ in hexagonal lattice for several values of the synergy rate $\beta$ as marked in the legend. 
The circles correspond to the critical values of $\alpha_c(\beta)$.}
\label{fig:hex}
\end{figure}

We start with analysis of the SIR process on a hexagonal lattice ($q=3$). 
The phase diagram is shown in Fig.~\ref{fig:hex}(a). 
The results of analytical model are shown by solid symbols. 
They were obtained by numerical solution of implicit Eq.~\eqref{eq:boundary_1} in the range $\beta \ge -\alpha/2$ (to the right of the dashed line), 
of Eqs.~\eqref{eq:T_mean_general}-\eqref{eq:P_n},~\eqref{eq:P_0}~\eqref{eq:T_mean_cube} and ~\eqref{eq:I_2_B0} in the range $-\alpha \le \beta < -\alpha/2$ (between the dotted and dashed lines) and of 
Eq.~\eqref{eq:alpha_c_hex_neg_beta} for $ \beta \le -\alpha $ with $T_c=1-2\sin(\pi/18)\simeq 0.653$~\cite{Sykes_1964}. 
The solution of these equations requires the knowledge of the function $s(\alpha, \beta)$. 
The probability $s(\alpha, \beta)$ can be calculated numerically by counting the relative number of 
non-synergistic attacks (successful or non-successful) from node $i$ to node $j$ for given values of $\alpha$ and 
$\beta$ and equating this to $(1-s(\alpha,\beta))^{q-1}$. 
The dependence of $s$ on $\alpha$ is shown in  Fig.~\ref{fig:hex}(b) for several values of $\beta$. 

The crossing points between the straight lines, $\alpha=-n\beta$ ($n=1,\ldots,q-1$), and $\alpha_c(\beta)$ in Fig.~\ref{fig:hex}(a) indicate the location of the kinks on the phase boundary and separate different synergy regimes. 
Comparison with the results of the exact numerical simulations (open symbols) confirms that the analytical model with a time-dependent environment gives reliable estimates for the phase boundary for not very strong positive synergy, i.e. in the range $\beta \alt 0.5$. 
For large values of $\beta$,  the critical value obtained by finite-size scaling tends to  
$\alpha_c \to 1.006\pm0.004$ for $\beta\to\infty$ while the estimate according to Eq.~\eqref{eq:boundary_1} is $\alpha_c \to 0.94 \pm 0.01$, meaning that the theoretical model slightly underestimates the invasion threshold.
As seen from Fig.~\ref{fig:hex}(a), the synergy effects in transmission rate do not change significantly the invasion threshold in the synergy-free case, 
$\alpha_c(\beta=0)\simeq 1.058$, leading to its variation in a relatively small range around $\alpha_c(\beta=0)$, 
$\alpha_c(\infty)<\alpha_c(\beta)<\alpha_c(-\infty)$ (where  
$\alpha_c(\pm\infty)=\lim_{\beta\to\pm\infty}\alpha_c(\beta)$), with 
$|\alpha_c(-\infty)-\alpha_c(\infty)|/\alpha_c(0)\sim 0.1 \ll 1$. 
Eq.~\eqref{eq:boundary_small_beta} suggests that such  an insignificant effect of synergy on invasion threshold in a hexagonal lattice can be related to its low coordination number, $q=3$\, and the small value of the probability $s$ that two or more infected hosts challenge a common neighbour during their infectious period. 
The function $s(\alpha,\beta)$ used for evaluation of $\alpha_c(\beta)$ is presented in  Fig.~\ref{fig:hex}(b) against $\alpha$ for several values of $\beta$. 
As can be seen, $s$ takes small values at the invasion threshold (indicated by an open circle for each value of $\beta$). 
In particular, $s_0=s(\alpha_0,0)\simeq 0.0655$ in the absence of synergy.
The small values of $s$ at the invasion threshold indicate that the majority of transmission events are associated with a single infecting node. 
This is ultimately associated with the poor connectivity of the hexagonal lattice that has a small coordination number and a relatively long path between neighbours of a susceptible recipient node. 
Indeed, the shortest path between two neighbours of a recipient host (excluding the path through that recipient) consists of four edges. 
Equivalent paths for square and triangular lattices are of length of two and one edges, respectively.

\begin{figure}[h] 
\centering
\begin{subfigure}[b]{0.4\textwidth}
\centering
{\includegraphics[clip=true,width=\textwidth]{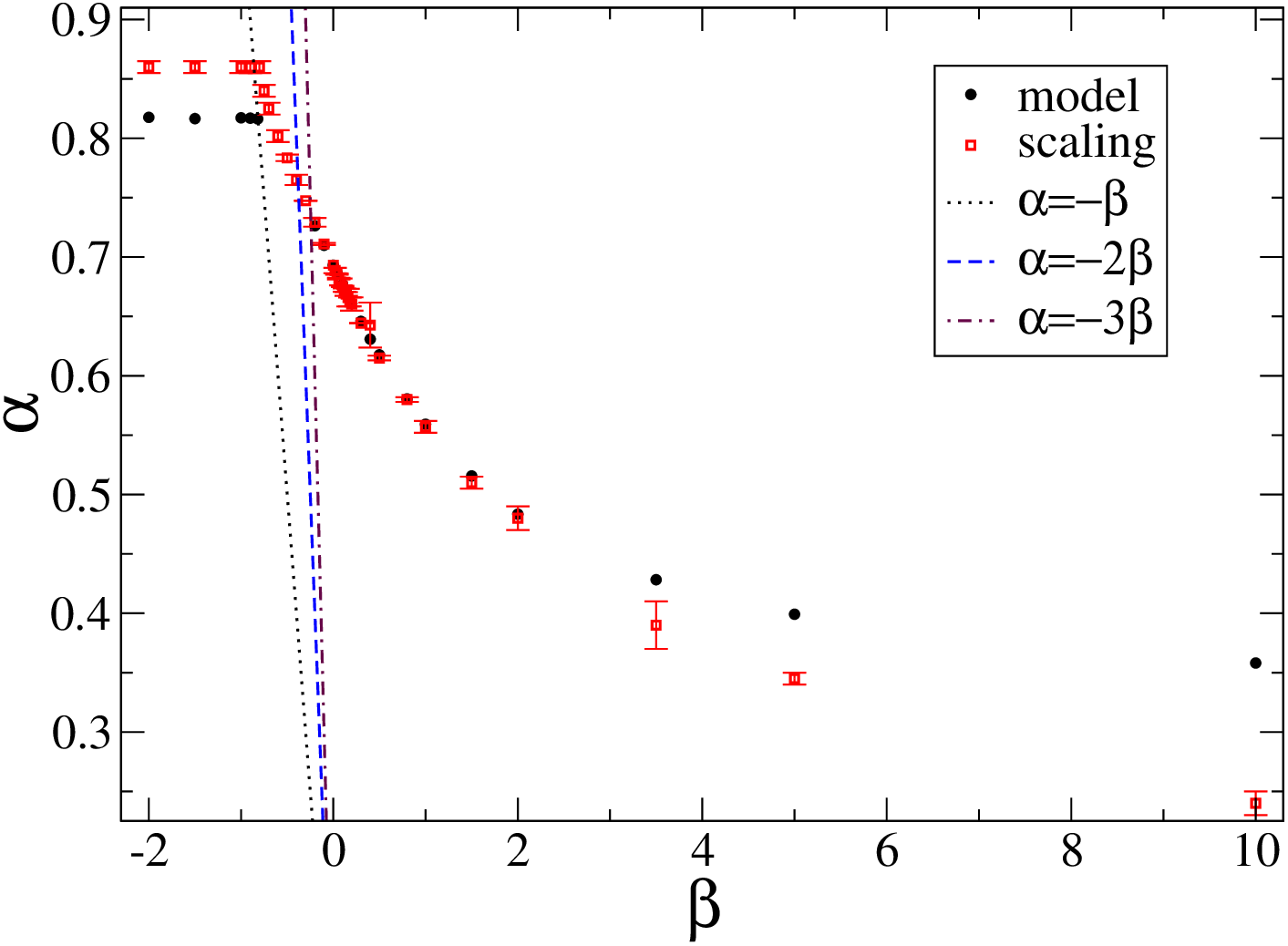}}   
\end{subfigure}
\quad
\begin{subfigure}[b]{0.4\textwidth}
\centering
{\includegraphics[clip=true,width=\textwidth]{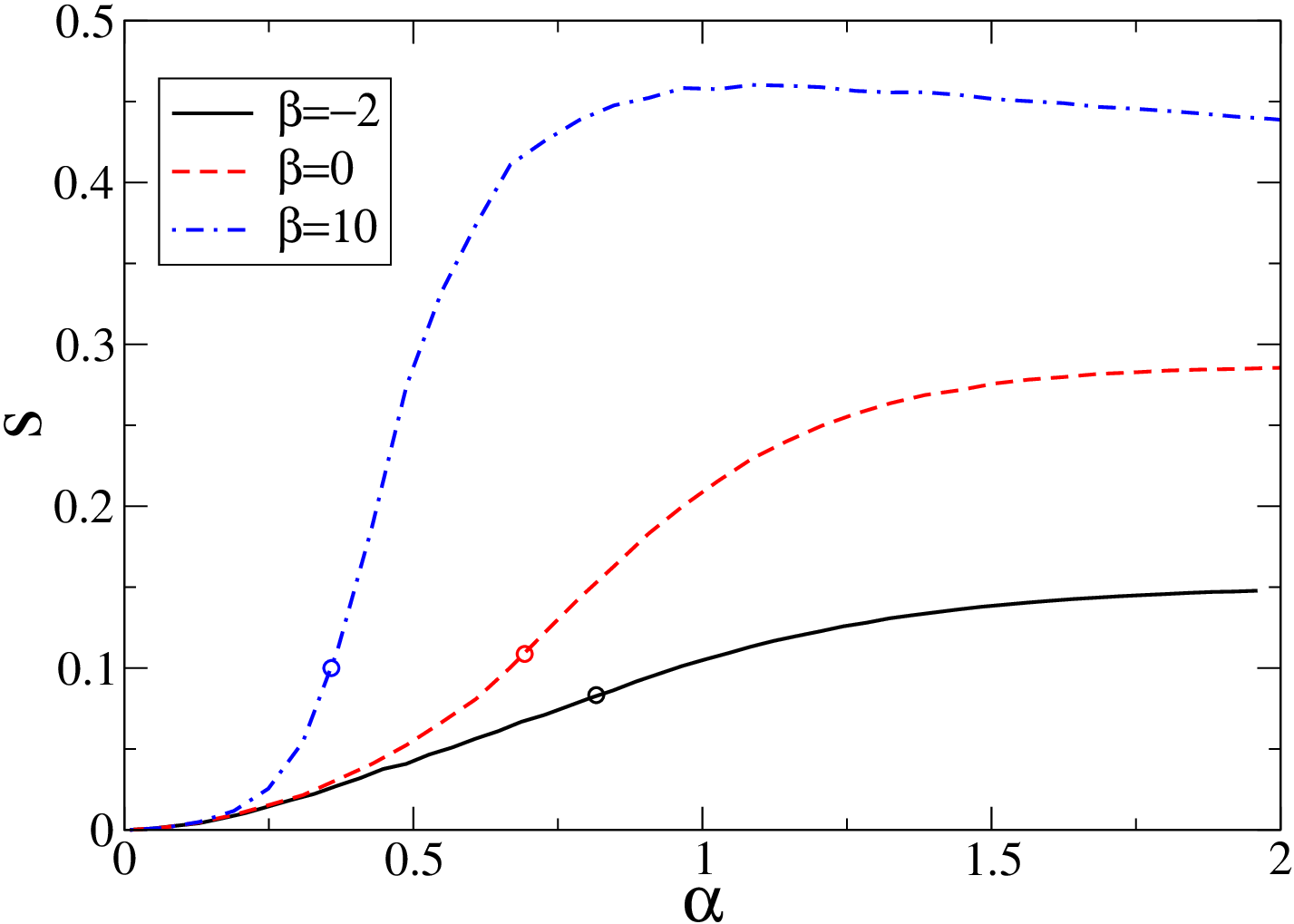}} 
\end{subfigure}
\caption{ (a) Phase diagram for the square lattice. 
The results of numerical finite-size scaling analysis are shown by open symbols.  The solid symbols correspond to solution of analytical model. The dotted and dashed lines correspond to $\alpha=-\beta$ and $\alpha=-2\beta$. \\
(b) Infection probability $s$ {\it vs} synergy-free transmission rate $ \alpha$ in the square lattice for several values of the synergy rate $\beta$ as marked in the legend. 
The circles correspond to the critical values $\alpha_c(\beta)$.}
\label{fig:square}
\end{figure}

In a square lattice, the effect of synergy is much more pronounced. 
The phase boundary shown in Fig.~\eqref{fig:square}(a) is similar in shape to that for a hexagonal lattice but covers a much wider interval of $\alpha$, 
i.e. $|\alpha_c(-\infty)-\alpha_c(\infty)|/\alpha_c(0) \sim 1$. 
The analytical model gives a  phase boundary (solid symbols) that reproduces  well 
the numerical data in the range, $-0.2 \alt \beta \alt 2$, slightly underestimating the invasion threshold $\alpha_c(-\infty)$ ($\alpha_c(-\infty)\simeq 0.86 \pm 0.005$  by finite-size scaling analysis 
and $\alpha_c(-\infty)\simeq 0.82 \pm 0.01$ from the model) and overestimating 
 $\alpha_c(\infty)$ ($\alpha_c(\infty)\simeq 0.13 \pm 0.05$ by scaling analysis 
and $\alpha_c(\infty)\simeq 0.34 \pm 0.01$ according to the model). 
The reason for stronger effect of synergy on invasion threshold in the square  lattice as compared to  hexagonal one is due to higher network connectivity (i.e. greater coordination number,  $q=4$,  and shorter path between neighbours of a recipient node). 
The better connectivity of the network leads to greater values of $s$ for the probability 
of infection of nearest neighbours of susceptible node both at the phase boundary, $\alpha_c(\beta) \sim 0.1$ (see the circles in Fig.~\ref{fig:square}(b)) and in the limit for large constructive synergy   $\alpha_c(\infty) \sim 0.4$ (see the plateau value for the dot-dashed curve in Fig.~\ref{fig:square}(b)), respectively.

\begin{figure}[h] 
\centering
\begin{subfigure}[b]{0.4\textwidth}
\centering
{\includegraphics[clip=true,width=\textwidth]{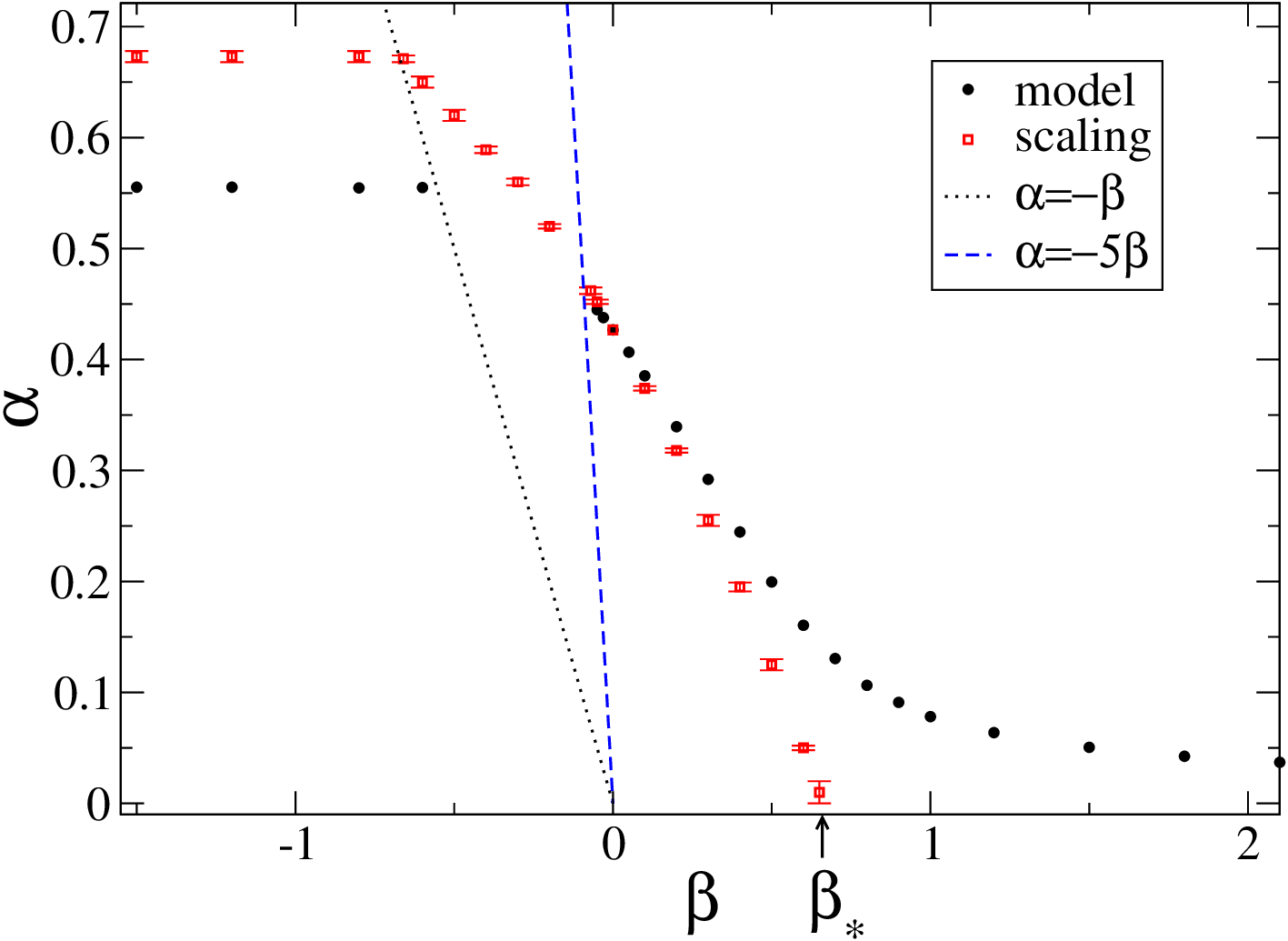}}   
\end{subfigure}
\quad
\begin{subfigure}[b]{0.4\textwidth}
\centering
{\includegraphics[clip=true,width=\textwidth]{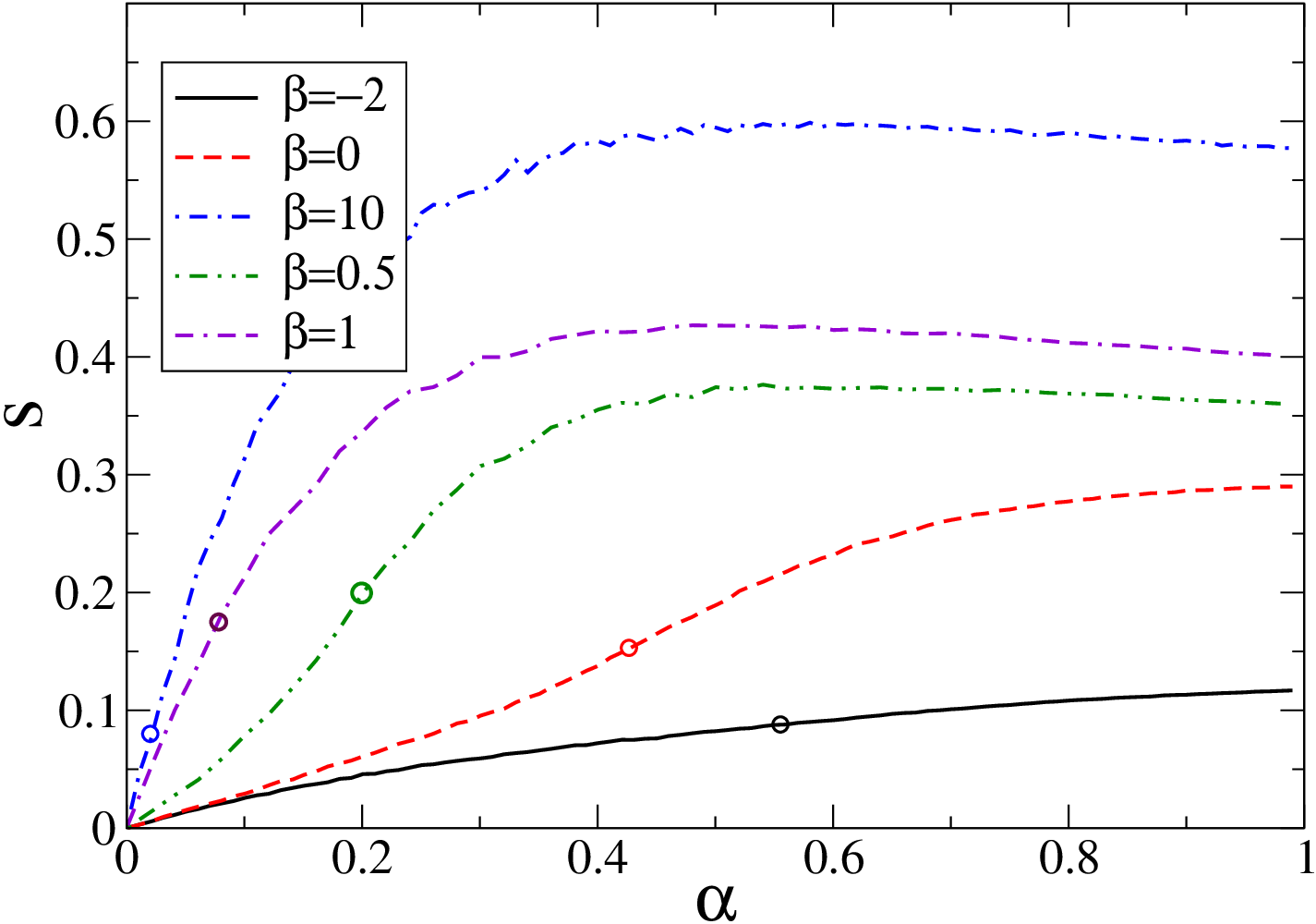}} 
\end{subfigure}
\caption{ (a) Phase diagram for triangular lattice. 
 The dotted and dashed lines correspond to $\alpha=-\beta$ and $\alpha=-5\beta$. The same symbols are used as in Fig.~\ref{fig:hex}. \\
(b) Infection probability $s$ {\it vs} synergy-free transmission rate $ \alpha$ in the triangular lattice for several values of the synergy rate $\beta$ as marked in the legend. 
The circles correspond to the critical values $\alpha_c(\beta)$. }
\label{fig:tr}
\end{figure}

A triangular lattice is the best connected one with coordination number $q=6$ and a path of length of one edge between neighbours of a recipient host. 
Consequently, the synergy effects are the most prominent for this lattice. 
The region corresponding to non-invasive epidemics in the $(\alpha,\beta)$ plane  is significantly reduced and an invasion can even occur at $\alpha\to 0$ if the synergy rate is
larger than a certain value $\beta_* \simeq 0.65$ 
(marked by the arrow in Fig.~\ref{fig:tr}(b)). 
The value of the critical synergy rate $\beta_*$, which is the solution of $\alpha_c(\beta_*)=0$, is slightly greater than the synergy-free critical rate, 
i.e. $\beta_*> \alpha_c(0)=-\ln(1-T_c) \simeq 0.427$ with the bond percolation threshold 
$ T_c = 2\sin(\pi/18)\simeq 0.347$. 
The reason for this is that synergistic attacks typically
last less time than the life-time of one of the infecting nodes because one of the attackers would move to the removed class and, at this point, the infection rate of the remaining node becomes zero. 
This is in contrast to the synergy-free case when the infection can last for 
the whole life-time of the infecting node. 

The theoretical model describes the phase-boundary reasonably well for $|\beta|\ll 1$ but fails to reproduce its crossing with the horizontal axis. 
Instead, the value of $\alpha$ asymptotically approaches the finite value 
$\alpha_c(\infty) \simeq 0.02$. 
The failure of the model could be related to the brake-down of the assumption of uncorrelated transmissibilities used both in Eq.~\eqref{eq:condition} and Eq.~\eqref{eq:T_mean_general}. 
Indeed, in highly connected triangular lattice, 
the node $i$ attacking node $j$ is linked directly to two nearest neighbours of node $j$ and thus it certainly attempts to transmit infection to them. 
This means that the infection times of these two nearest neighbours of $i$ are correlated with the infection time of node $i$, which is not taken into account within the simple model for time-dependent environment.

\section{\label{sec:Synchrony}Effects of synchrony in transmission}

In previous sections, it has been assumed that elementary transmission in the absence of synergy obeys asynchronous dynamics with constant rate, $\lambda(t)=\alpha$. 
In this case, the good agreement between exact analytical results and numerical simulations justifies the uniform distribution for infection times $t_{j'}$ given by Eq.~\eqref{eq:rho_uniform}. 
The characteristics of the final state of non-synergistic SIR epidemics are known to be independent of the functional form of $\lambda(t)$, provided it only depends on time since infection of the donor host~\cite{Ludwig_MBiosc1975,PerezReche_Interface2012}. 
In other words, the final state of SIR epidemics is statistically identical irrespective of whether transmission of infection obeys a homogeneous or inhomogeneous Poisson process. 
Such interesting property does not hold in general for synergistic transmission. 
In order to analyse the interplay between time-dependent elementary transmission rate and synergy, we first consider discrete-time synchronous transmission dynamics and then investigate the effect of weak deviations from the purely asynchronous dynamics with synergy studied in previous sections. 
The analysis in this section is restricted to the regime $\beta \geq  - \alpha/(q-1)$.

Discrete-time transmission occurs instantaneously at the end of the life time of the transmitting node, $t=\tau=1$, and thus the elementary transmission rate has a $\delta$-functional form, $\alpha= \delta(t-1)$.
In the absence of synergy, this corresponds to the discrete-time SIR model or Reed-Frost model~\cite{Ludwig_MBiosc1975,PerezReche_Interface2012} in which all infected hosts simultaneously transmit infection with probability $T=1-\exp(-\alpha)$ (cf. Eq.~\eqref{eq:T_t}) at the end of their infectious period (i.e. at time $t=1$ since infection). 
The discrete-time nature of transmission implies that all hosts concurrently challenging a common recipient neighbour become infected simultaneously, meaning that,
\begin{equation} 
\rho(t_{j'})=s\delta(t_{j'})~.
\label{eq:rho_discrete}
\end{equation}
This also means that synchronous transmission leads to time-independent environment during transmission -- the assumption used in the synergy toy model of 
Ref.~\cite{Perez_Reche_2011:PRL}.
Indeed, use of $\rho(t_{j'})=s\delta(t_{j'})$ in Eq.~\eqref{eq:fbeta_def} gives  $f(\beta)=e^{-\beta}$ which coincides with the function $f(\beta)$ obtained in Sec.~\ref{sec:sol_1} for the time-independent environment model of Ref.~\cite{Perez_Reche_2011:PRL}. 
Consequently, the synergy-induced deviation of the critical rate $\alpha_c$ from the non-synergistic value $\alpha_0$ is twice larger for synchronous transmission than it is for asynchronous epidemics (see Eq.~\eqref{eq:boundary_small_beta}), i.e. synchrony has a significant impact on synergistic epidemics. 

To further investigate the role of synchrony in transmission, we now consider an elementary transmission rate $\alpha A(t)$. 
The function $A(t)$ represents a small deviation from the homogeneous Poisson process characterised by $A(t)=1$. 
Without lost of generality, we restrict the analysis to functions $A(t)$ satisfying $\int_0^1A(t)\text{d}t=1$ in order for Eq.~\eqref{eq:r}  and thus all the equations derived in Sec.~\ref{sec:sol_1} for general $\rho(t_{j'})$ to hold. 
We assume that a small deviation from the homogeneous Poisson transmission process results in a slightly non-uniform probability density function, $\rho(t_{j'})$, that we consider to be piece-wise linear to first approximation:
\begin{equation}
\label{eq:rho_sigma}
\rho(t_{j'})=\frac{\sigma}{2}\left(-\left|t_{j'}\right|+\frac{1}{2} \right)+\frac{s}{2}~,
\end{equation}
for $t_{j'}\in[-1,1]$ and $\rho(t_{j'})=0$ otherwise, 
where $s/\sigma \geq 1/2$ to ensure that $\rho$ is non-negative.
The function $\sigma=\sigma(\alpha,\beta)$ quantifies the synchrony between attacker $i$ infected at time $t=0$ and attacker $j'$ infected at time $t_{j'}$. 
The uniform distribution given by Eq.~\eqref{eq:rho_uniform} is recovered for $\sigma=0$. 
Positive values of $\sigma$ correspond to cases in which the time of infection of $j'$ is typically closer to that of node $i$ than for pure asynchronous dynamics corresponding to Eq.~\eqref{eq:rho_uniform}. 
For $\sigma >0$, the function $\rho(t_{j'})$ has a maximum at $t_{j'}=0$ and thus is qualitatively similar to the $\delta$-functional distribution for 
fully synchronous transmission (see Eq.~\eqref{eq:rho_discrete}).
In contrast, asynchrony is promoted for negative values of $\sigma$.

The function $f(\beta)$ (cf. Eq.~\eqref{eq:fbeta_def}) corresponding to the probability density function given by Eq.~\eqref{eq:rho_sigma} is,
\begin{equation}
f(\beta)=-\frac{\sigma}{s}\left(f_1(\beta) - \frac{f_0(\beta)}{2} \right)+f_0(\beta)~,
\end{equation}
with $f_0(\beta)$ given by Eq.~\eqref{eq:fbeta_uniform} and
\begin{equation}
\label{eq:fbeta_1}
f_1(\beta)=\frac{1}{\beta^2}\left(\beta-1+e^{-\beta}\right)~.
\end{equation}

For $\beta \rightarrow 0$, the expansion given by Eq.~\eqref{eq:boundary_small_beta} results in
\begin{equation}
\label{eq:alphac_linear_sigma}
\alpha_\text{c}(\beta)=\alpha_0-\frac{q-1}{2}\left(s_0+\frac{\sigma_0}{6} \right)\beta~,
\end{equation}
where $\sigma_0$ is the function $\sigma$ evaluated at the non-synergistic invasion threshold, i.e.  $\sigma_0=\sigma(\alpha_0,0)$.

Eq.~\eqref{eq:alphac_linear_sigma} shows that, as expected, (i) synchrony only plays a role when $\beta \neq 0$ and (ii) the effect of synergy on the invasion threshold is more significant for positive values of $\sigma_0$ corresponding to some degree of synchrony. 
The fact that the first-order correction to $\alpha_\text{c}$ is proportional to the sum of $\sigma_0$ and $s_0$ (rather than their product) demonstrates that the effect of synchrony between concurrent attackers (quantified by $\sigma_0$) is of the same order as the probability $s_0$ of that their infectious periods overlap. 

\section{\label{sec:Conclusions}Conclusions}

To conclude, 
we have developed and solved an analytical model for the SIR process on a lattice with synergy effects caused by interactions between nodes that challenge simultaneously a given recipient node.
Our main findings are the following. 
(i) For lattices with low coordination number (e.g. a hexagonal lattice),  
the synergy effects are insignificant and the critical value of transmission rate practically does not depend on synergy-dependent rate.  
(ii) In contrast, the lattices with high coordination number (e.g. a triangular lattice) become less resilient due to synergy effects and even for very small values of transmission rate the SIR process can be  invasive for sufficiently large values of the synergy rate.     
(iii) We have developed a theoretical framework for description of the SIR epidemics with synergy incorporating significant  effects of a time-dependent 
environment. 
The analytical solution of the model allows the phase boundary for transmission rates to be found and analysed. 
The theory gives reliable estimates for weak synergy in case of all studied 2d topologies. 
It also gives very good estimates for the phase boundary in lattices with relatively low connectivity (e.g. hexagonal and square) but fails in highly connected lattices (e.g. triangular) due to strong correlation effects in transmissibilities. 
It is demonstrated that accounting for the evolution of the neighbourhood of recipient hosts is crucial to properly describe synergistic spreading processes.
The role of synchrony between infected hosts challenging a common recipient has also been analysed in terms of the proposed model. 
Results confirm the fact that synchrony does not play any role for non-synergistic SIR epidemics but it plays a significant role for synergistic transmission.

\appendix
\section{\label{app:A}Reduction formula for $I_n$ in the regime, 
$\beta < -\alpha$}

In this Appendix, the reduction formula for integrals $I_n$ is derived. 

We start with the simplest cases of $n=1$ corresponding to situations in which the recipient node is challenged by two infected neighbours. 
In the presence of a single $(n=1)$ infected neighbour (in addition to attacker), the 
expression for transmission rate has the following form, 
\begin{equation}  
r(t_1)=\left\{ 
\begin{array}{lllll}
& \int_0^{t_1}\alpha \text{d}t'= \alpha t_1 ~~ & \text{if}& ~~ &t_1 \in [0,1]
\\
& \int_{t_1+1}^{1}\alpha \text{d}t'=-\alpha t_1~~  & \text{if}& ~~ &t_1 \in [-1,0) 
\end{array}
\right. 
\label{eq:r_t1} 
\end{equation} 
i.e., $r(t_1)= \alpha |t_1|$. Here, $t_1$ is the time of infection of the second attacker.

The integral $I_1$ is therefore,
\begin{equation}  
I_1 = I_1^{(-)} + I_1^{(+)}= 
\int\limits_{-1}^0\left(1-e^{\alpha t_1} \right)\text{d}t_1
+ 
\int\limits_{0}^1\left(1-e^{-\alpha t_1} \right)\text{d}t_1=
 2\left(1-\frac{I_0}{\alpha}\right)
~, 
\label{eq:I_1} 
\end{equation} 
where we used Eq.~\eqref{eq:P_0}.
The factor $2 (= n+1)$ in Eq.~\eqref{eq:I_1} refers to the number of distinct way of placing $t_1\in[-1,1]$, i.e. (i) $t_1\in[0,1]$ when $r(t_1)=\alpha t_1$ 
and (ii) $t_1\in[-1,0)$ when $r(t_1)=-\alpha t_1$. 

\begin{figure}[ht] 
\begin{center} 
{\includegraphics[clip=true,width=12cm]{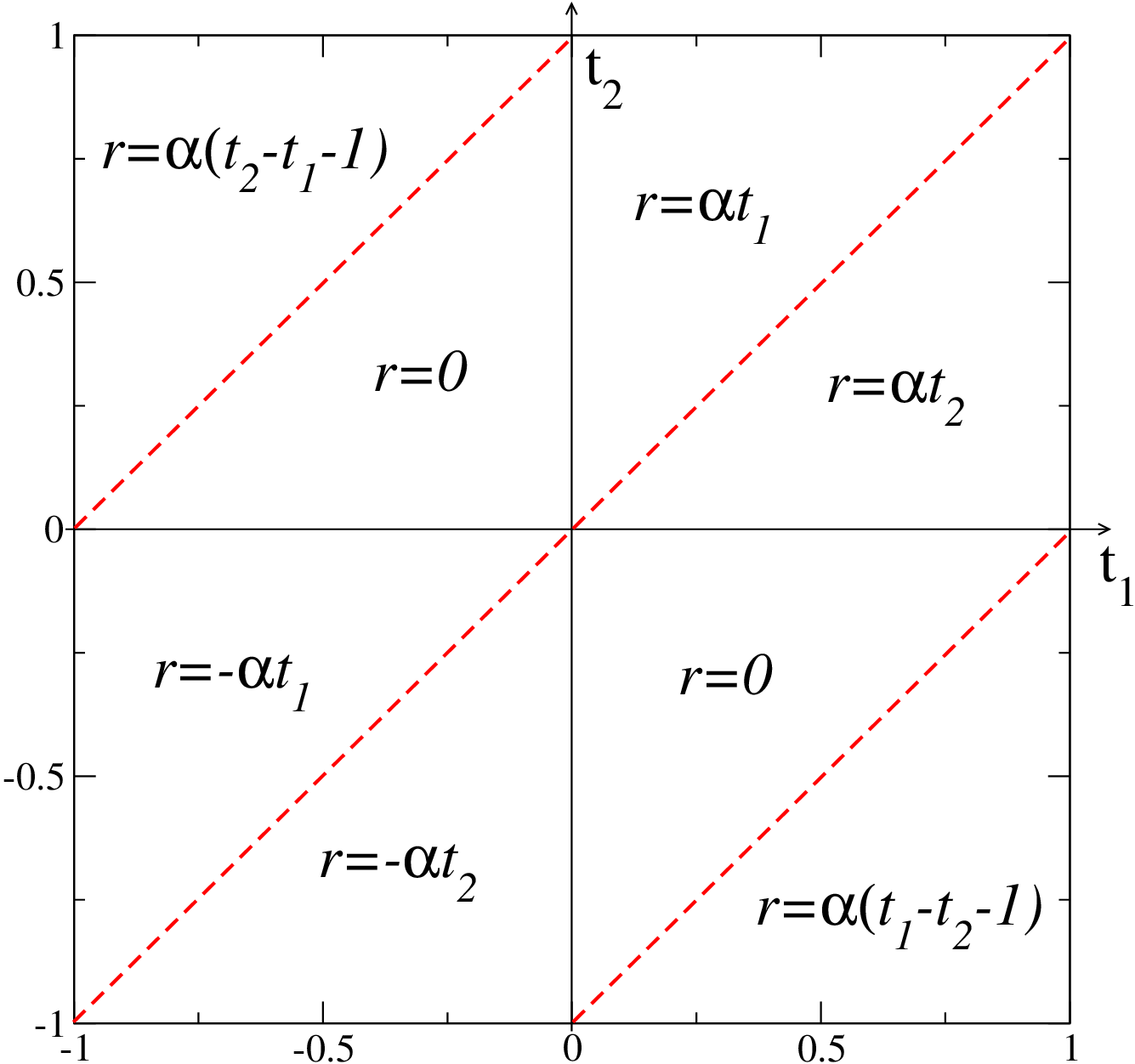}} 
\end{center} 
\caption{\label{fig:t1t2} The expressions for $r(t_1,t_2)$ in different 
regions of the square $C_2$ in the  regime $\beta \le - \alpha$.
} 
\end{figure}

In the case of coordination number $q=3$ and two possible nearest neighbours 
which might exhibit synergy effects during infection, the expression for $r(t_1,t_2)$ depends on mutual location of infection times $t_1$ and $t_2$ in the interval $[-1,1]$. 
If both neighbours influence the transmission of infection, the expression for $r(t_1,t_2)$ can have four possible functional forms in the square 
$C_2$ (where $C_n$ stands for the interior domain of n-dimensional cube with sides of length $2$, see the definition after Eq.~\eqref{eq:P_n})   
centred at the origin of the $(t_1,t_2)$ parameter space (see Fig.~\ref{fig:t1t2}). 
This is a consequence of the mirror-reflection symmetry of the problem about bisector $t_2=t_1$ with simultaneous swap $t_1 \leftrightarrows t_2$.
More precisely, $r(t_1,t_2)=0$ in two triangular regions, $(t_1\in[0,1],t_2\in [t_1-1,0])$ and 
$(t_1\in[-1,0),t_2\in [0,t_1+1])$, reflecting strong destructive interference from the neighbours resulting in zero transmission rate during the whole life-time of the attacker. 
In other regions, $r(t_1,t_2)=\alpha t_1$ for $(t_1\in[0,1],t_2\in (0,t_1+1])$,  $r(t_1,t_2)=\alpha (t_1-t_2-1)$ for $(t_1\in[0,1],t_2\in [-1,t_1-1))$ and   $r(t_1,t_2)=- \alpha t_2$ for $(t_1\in[-1,0),t_2\in [-1,t_1])$. 
With these rates, the expression for $I_2$ reads, 
\begin{eqnarray}  
&&I_2 = I_2^{(-)}+I_2^{(+)}=\int\limits_{-1}^0 \text{d}t_1 \int\limits_{-1}^1  \text{d}t_2
\left(1-e^{-r(t_1,t_2)} \right)+\int\limits_{0}^1 \text{d}t_1 \int\limits_{-1}^1  \text{d}t_2
\left(1-e^{-r(t_1,t_2)} \right)
\nonumber \\
&=& 
\int\limits_{-1}^0 \text{d}t_1\int\limits_{t_1}^0 \text{d}t_2 
\left(1-e^{\alpha t_2} \right) 
+
\int\limits_{-1}^0 \text{d}t_2\int\limits_{t_2}^0 \text{d}t_1 
\left(1-e^{\alpha t_1} \right) 
+
\int\limits_{-1}^0 \text{d}t_1 \int\limits_{t_1+1}^1 \text{d}t_2 
\left(1-e^{-\alpha (t_2-t_1-1)} \right) 
\nonumber \\
&+&
\int\limits_{0}^1 \text{d}t_2\int\limits_{0}^{t_2} \text{d}t_1 
\left(1-e^{-\alpha t_1} \right) 
+
\int\limits_{0}^1 \text{d}t_1\int\limits_{0}^{t_1} \text{d}t_2 
\left(1-e^{-\alpha t_2} \right) 
+
\int\limits_{0}^1 \text{d}t_1 \int\limits_{-1}^{t_1-1} \text{d}t_2 
\left(1-e^{-\alpha (t_1-t_2-1)} \right)~. \nonumber  \\
\label{eq:I_2} 
\end{eqnarray} 
All the contributions to $I_2^{(\pm)}$ are equal to each other and 
\begin{eqnarray}  
 I_2^{(\pm)}= 3\left(\frac{1}{2} \mp \frac{1}{\alpha} 
\int\limits_{0}^{\pm 1}\left(1-e^{\mp \alpha t_1} \right)\text{d}t_1 \right)=
3\left(\frac{1}{2}-\frac{1}{\alpha} 
I_1^{(\pm)} \right)
~.
\label{eq:I_2_minus} 
\end{eqnarray} 
The sum of these contributions results in the expression for $I_2$ in terms of 
$I_1$, 
\begin{eqnarray}  
 I_2= 
3\left(1 -\frac{I_1}{\alpha} 
 \right)
~.
\label{eq:I_2_sum} 
\end{eqnarray} 

For greater values of $n>2$, the expressions for $I_n$ can be derived in a recursive way resulting in the following relation,  
\begin{eqnarray}  
 I_n= 
(n+1)\left(1 -\frac{I_{n-1}}{\alpha} 
 \right)
~, ~~~n\ge 1~,  
\label{eq:I_n} 
\end{eqnarray} 
used in Sec.~\ref{sec:sol_2}. 
In the limiting case of zero infection rate, $\alpha\to 0$, the integrals $I_n$ tend to zero, and $\langle T \rangle\to 0$ as expected.

\section{\label{app:B}Mean transmissibility in the regime 
$-\alpha \le \beta < -\alpha/(q-1)$ for $q=3$}

In this Appendix, we present the derivation for integrals $I_n$ in the intermediate regime with $-\alpha \le \beta < -\alpha/(q-1)$
for $q=3$ (hexagonal lattice). 
\begin{figure}[!] 
\begin{center} 
{\includegraphics[clip=true,width=12cm]{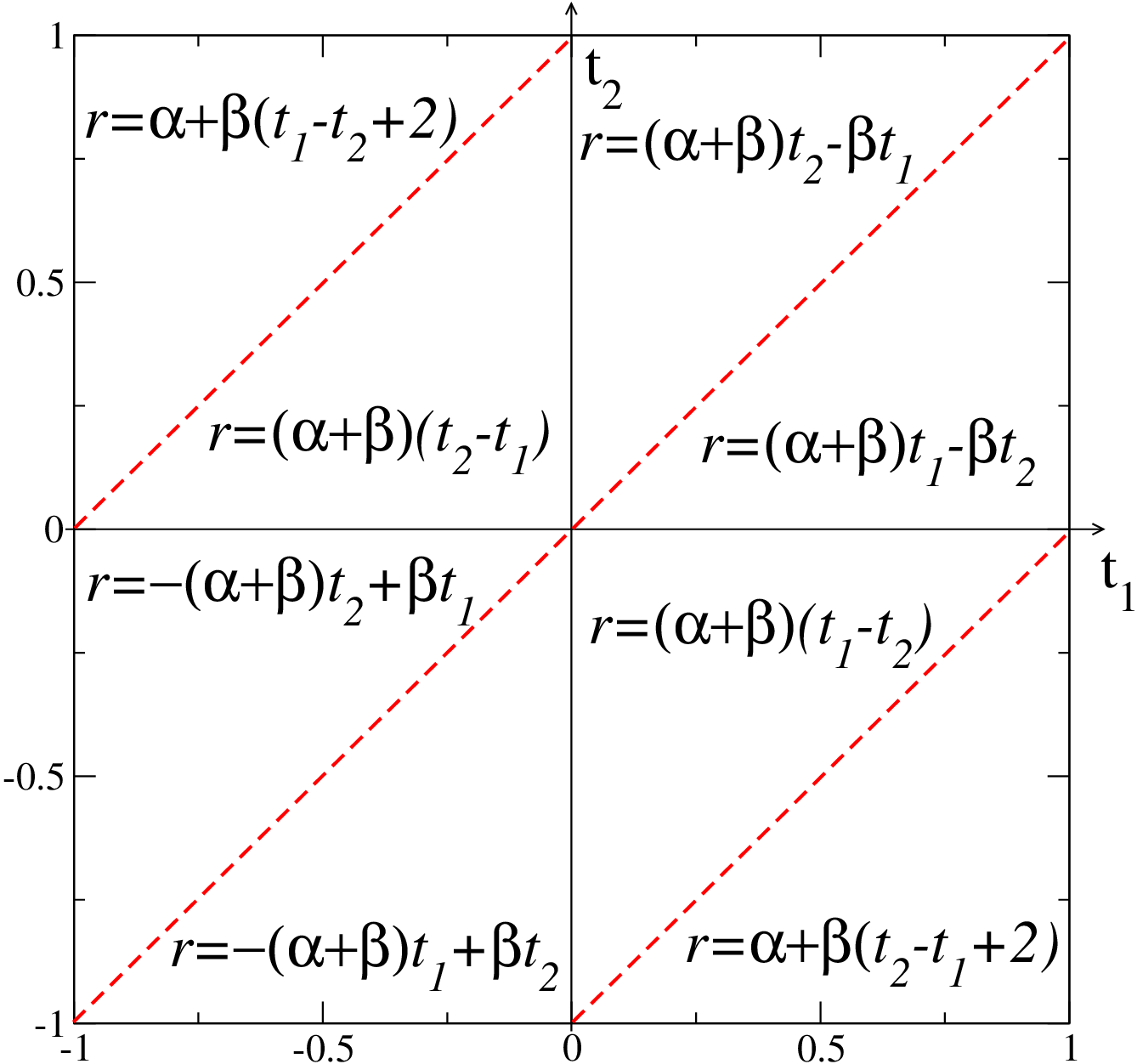}} 
\end{center} 
\caption{\label{fig:t1t2_2} 
The expressions for $r(t_1,t_2)$ in different 
regions of the square $C_2$ in the  regime 
 $-\alpha \le \beta < -\alpha/2$. 
} 
\end{figure}

In this regime, 
the transmission rate can be non-zero in the presence of none or single infected neighbours of $j$ (excluding node $i$ attacking node $j$). 
Consequently, the expressions for $I_0$ and $I_1$ coincide with those obtained for the $\beta \ge -\alpha/(q-1) $ regime and are given by Eqs.~\eqref{eq:P_0} and ~\eqref{eq:T_mean_cube}. 
However, the value of $I_2$ is different in this regime because the transmission rates becomes zero in the presence of two infected nearest neighbours of $j$. 
The transmission rate, $r(t_1,t_2)$, depends on the mutual location of infection times $t_1$ and $t_2$ within the square, $t_{j'}\in[-1,1]$ ($j'=1,2$), and has different functional form  in 8 triangular regions (see Fig.~\ref{fig:t1t2_2}). 
The expression for $I_2$ reads, 
\begin{eqnarray}  
I_2 = I_2^{(-)}+I_2^{(+)}=\int\limits_{-1}^0 \text{d}t_1 \int\limits_{-1}^1  \text{d}t_2
\left(1-e^{-r(t_1,t_2)} \right)+\int\limits_{0}^1 \text{d}t_1 \int\limits_{-1}^1  \text{d}t_2
\left(1-e^{-r(t_1,t_2)} \right)  
~,
\label{eq:I_2_B} 
\end{eqnarray} 
with $I_2^{(-)}=I_2^{(+)}$ by symmetry. 
The area of integration for $I_2^{(+)}=\sum_{k=1}^4I_{2,k}^{(+)}$ can be split into $4$ triangles (see Fig.~\ref{fig:t1t2_2}), i.e. 
\begin{eqnarray}  
I_2 =2\sum_{k=1}^4I_{2,k}^{(+)}
~,
\label{eq:I_2_B0} 
\end{eqnarray} 
with the following contributions, 
\begin{eqnarray} 
I_{2,1}^{(+)}=
\int\limits_0^1 \text{d}\, t_{1} \int\limits^1_{t_1} \text{d}\, t_{2} 
\left( 1- e^{-(\alpha+\beta)t_2 + \beta t_1 }\right)
=
\frac{1}{2} - \frac{1}{\beta} \frac{\left(1-e^{-\alpha}\right)}{\alpha} 
+ \frac{1}{\beta} \frac{\left(1-e^{-(\alpha+\beta)}\right)}{\alpha+\beta}
~,
\label{eq:I_2_B1} 
\end{eqnarray} 
\begin{eqnarray} 
I_{2,2}^{(+)}= 
\int\limits_0^1 \text{d}\, t_{1} \int\limits_0^{t_1} \text{d}\, t_{2} 
\left( 1- e^{-(\alpha+\beta)t_1 + \beta t_2 }\right)=I_{2,1}^{(+)}
~,
\label{eq:I_2_B2} 
\end{eqnarray} 
\begin{eqnarray} 
I_{2,3}^{(+)}
= 
\int\limits_0^1 \text{d}\, t_{1} \int\limits_{t_1-1}^{0} \text{d}\, t_{2} 
\left( 1- e^{ -(\alpha+ \beta)(t_1-t_2) }\right)
=
\frac{1}{2} + \frac{e^{-(\alpha+\beta)}}{\alpha+\beta}  - 
\frac{1-e^{-(\alpha+\beta)}}{(\alpha+\beta)^2}
~,
\label{eq:I_2_B3}
\end{eqnarray} 
and the last contribution which coincides with that obtained for regime $\beta \ge -\alpha/2 $, i.e. 
\begin{eqnarray} 
I_{2,4}^{(+)}
=  
\int\limits_0^1 \text{d}\, t_{1} \int\limits_{-1}^{t_1-1} \text{d}\, t_{2} 
\left( 1-  e^{ -\alpha- \beta(t_2-t_1+2) }\right)
=\frac{1}{2} - \frac{e^{-(\alpha+\beta)}}{\beta}
\left(    
\frac{e^{\beta}-1}{\beta} - 1
\right)
~. 
\label{eq:I_2_B4}
\end{eqnarray} 
The final expression for transmissibility is given by Eqs.~\eqref{eq:T_mean_general}-\eqref{eq:P_n} with $I_0$, $I_1$ and $I_2$ 
given by ~Eqs.~\eqref{eq:P_0}, ~\eqref{eq:T_mean_cube} and ~\eqref{eq:I_2_B0}. 
A similar analysis in regimes $-\alpha/(n-1) \le \beta < -\alpha/n$, for $n=2,\ldots, q-1$  can be done for $q>3$.

\end{document}